# Poly(acrylic acid)-coated iron oxide nanoparticles : quantitative evaluation of the coating properties and applications for the removal of a pollutant dye

**J. Fresnais[1]\*, M. Yan[2], J. Courtois[2], T. Bostelmann[1], A. Bée[1] and J.-F. Berret[2]\***

[1] *Physicochimie des Electrolytes, Colloïdes et Sciences Analytiques (PECSA) UMR 7195 CNRS-UPMC- ESPCI, 4 place Jussieu, 75252 Paris Cedex 05*

[2] *Matière et Systèmes Complexes, UMR 7057 CNRS Université Denis Diderot Paris-VII, Bâtiment Condorcet, 10 rue Alice Domon et Léonie Duquet, 75205 Paris (France)*

**Abstract:** In this work, 6 to 12 nm iron oxide nanoparticles were synthesized and coated with poly(acrylic acid) chains of molecular weight 2100 g mol$^{-1}$. Based on a quantitative evaluation of the dispersions, the bare and coated particles were thoroughly characterized. The number densities of polymers adsorbed at the particle surface and of available chargeable groups were found to be $1.9 \pm 0.3$ nm$^{-2}$ and $26 \pm 4$ nm$^{-2}$, respectively. Occurring via a multi-site binding mechanism, the electrostatic coupling leads to a solid and resilient anchoring of the chains. To assess the efficacy of the particles for pollutant remediation, the adsorption isotherm of methylene blue molecules, a model of pollutant, was determined. The excellent agreement between the predicted and measured amounts of adsorbed dyes suggests that most carboxylates participate to the complexation and adsorption mechanisms. An adsorption of 830 mg g$^{-1}$ was obtained. This quantity compares well with the highest values available for this dye.

**Keywords:** magnetic nanoparticles – polyelectrolytes – electrostatic complexation – methylene blue – adsorption isotherm – pollutant remediation

\*Corresponding authors   jean-francois.berret@univ-paris-diderot.fr
jerome.fresnais@upmc.fr

## 1.   Introduction

Nanotechnology's ability to shape matter at the scale of the nanometer has opened the door to new generations of applications in material science and nanomedicine. Engineered inorganic nanoparticles are key actors of this strategy. The particles are made of metallic or rare earth atoms and display fascinating size-related optical or magnetic properties [1]. One of the most appealing applications to emerge in the recent years is the use of iron containing particles for cleaning up contaminants in groundwater, soil and sediments [2, 3]. The benefit of using magnetic particles and composites (as compared to regular adsorbents [4]) is that the particles can be separated by the application of a magnetic field. Pollutants can thus be isolated and eliminated afterwards by simple and affordable techniques.

Many strategies devoted to the magnetic removal of organic, dye and metallic pollutants from wastewater have been implemented. These strategies are based on the fabrication of novel organic/inorganic nanocomposites in which magnetic nanoparticles are incorporated and provide the stimuli-responsive properties [3, 5-16]. The most conspicuous and recent examples are those of the magnetite/reduced graphene oxide composites [13] or of the carbon/cobalt ferrite alginate beads [17]. These newly synthesized nanomaterials are tested for the decontamination of model dyes and pollutant, such as methylene blue, rhodamine B or methyl orange [16, 18, 19]. For the remediation of heavy metallic ions ($Hg^{2+}$, $Pb^{2+}$, $As^{3+}$) from ground water and soils, alternative composites were generated. They comprised magnetic mesoporous silica beads [3, 16, 20, 21], hydrogels [22] or multiwall carbon nanotubes/iron oxide adsorbents [14, 23, 24]. Thanks to their increased specific surface, microporous beads and hydrogels [7, 15, 25] exploit electrostatic and non-specific adsorption to enhance the binding efficiency with low molecular weight contaminants. Because of their porous structures however, the later materials are characterized by long diffusion-reaction/adsorption times, and can be inappropriate for some applications.

Although the nanomaterials mentioned previously possess true potential for pollutant remediation, their physico-chemical properties are generally not studied in detail. Most reports focus on the performances and added values of these new adsorbents. These performances express in terms of adsorption isotherm or in terms mass of pollutant adsorbed per gram of materials [5, 7, 11-16, 22, 26, 27]. This latter quantity varies typically between 1 and 1000 mg/g for a wide variety of systems, including activated carbon, agricultural and industrial solid wastes and natural nanomaterials [4]. Very few papers in contrast investigate *i)* the colloidal stability of the nanomaterials in simple or in complex water-borne solvents and *ii)* the efficiency of the adsorption properties, *i.e.* the ratio between the observed and the nominal capacity to bind small molecules [28, 29]. These issues are crucial because they eventually determine the range of applications in natural ecosystems. If the adsorbents aggregate, they will be less efficient because of a reduction of specific surface. This is true for organics, and also for inorganic nanoparticles that are prone to destabilize through van der Waals attractive interactions. The knowledge of the adsorption efficacy is important since it decides *in fine* on the dose of particles to be used.

A broad range of techniques in chemistry and physical chemistry has been developed for the stabilization of inorganic nanoparticles [30]. Among these methods are the adsorption of charged ligands or stabilizers on their surfaces, the layer-by-layer (LbL) deposition of polyelectrolytes and the surface-initiated polymerization. Other approaches focused on the encapsulation of the particles in amphiphilic block copolymer micelles or in LbL capsules and surfaces. In the present paper, iron oxide with diameter 6 to 12 nm were synthesized and coated with poly(acrylic acid) of molecular weight 2100 g mol$^{-1}$ using the precipitation-redispersion principle. Combining scattering, thermogravimetry and titration experiments on three types of nano-sized iron oxides, a complete description of the particles and of their coating adlayers was realized. Free carboxylate functions attached to the particles are moreover excellent candidates for positively charged pollutant complexation, including dye molecules. With iron oxide as magnetic cores, the separation of the pollutant/nanoparticles complexes is achieved using an external magnetic field. The present work show that poly(acrylic acid) coated iron oxide nanoparticles are excellent adsorbents for methylene blue, with saturation adsorptions as high as 830 mg g$^{-1}$.

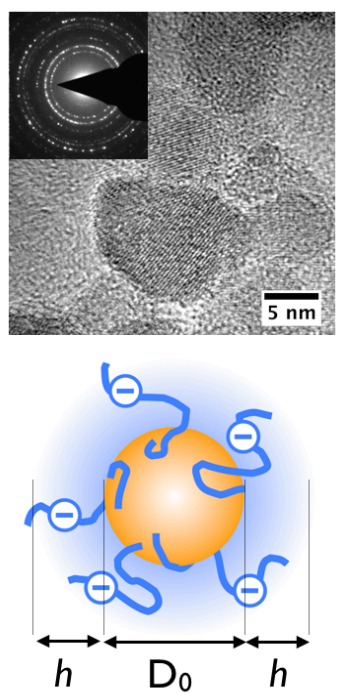

**Figure 1** : a) Transmission electron microscopy of iron oxide $\gamma$-Fe$_2$O$_3$. Inset: Bragg scattering rings obtained by electron beam micro-diffraction and identifying the structure of maghemite (Supporting Information, Fig. SI-2). b) Schematic representation of PAA$_{2K}$–$\gamma$-Fe$_2$O$_3$ nanoparticles.

## 2. Experimental

Iron oxide nanoparticles with bulk mass density $\rho$ = 5100 kg m$^{-3}$ were synthesized according to the Massart method [31] by alkaline co-precipitation of iron(II) and iron(III) salts and oxidation of the magnetite (Fe$_3$O$_4$) into maghemite ($\gamma$-Fe$_2$O$_3$) [32]. At *pH* 1.8, the bare particles are positively charged, with nitrate counterions adsorbed on their surfaces. The resulting inter-particle interactions are repulsive and impart an excellent colloidal stability to the dispersion (> years). The magnetic dispersions were characterized by vibrating sample magnetometry (VSM), by transmission electron microscopy (TEM) and by static and dynamic light scattering (SLS and DLS, respectively). These experiments allowed inferring the size distributions of the physical, magnetic and hydrodynamic diameters, noted $D_0^{TEM}$, $D_0^{VSM}$ and $D_H$ respectively. Determined from VSM measurements, the specific magnetization of the particles was found to be 3.5×10$^5$ A m$^{-1}$, *i.e.* slightly lower than that of bulk γ-Fe$_2$O$_3$ (4.5×10$^5$ A m$^{-1}$). A high-resolution TEM image and an electron beam micro-diffraction pattern (inset) of the particles are shown in Fig. 1a. For the present study, three batches of γ-Fe$_2$O$_3$ nanoparticles of median diameter $D_0^{VSM}$ = 6.7, 8.3 and 10.7 nm were synthesized. Table I lists the characteristics of the three batches in terms of diameters and of polydispersities. The polydispersity was defined as the ratio between standard deviation and average diameter. To improve their stability at neutral *pH*, the particles were coated with poly(acrylic acid) using the Precipitation-Dispersion protocol described earlier [33]. The molecular weight of the polymer was $M_W^{Pol}$ = 2100 g mol$^{-1}$ (in its sodium salt form), corresponding to 22 carboxylate monomers (polydispersity 1.7). When the iron oxide dispersion was mixed with that of the polymers, the solution underwent an instantaneous and macroscopic precipitation. The phase separation resulted from the multisite adsorptions *via* H-bonding of the uncharged acrylic acid moieties on the particle surfaces. As the *pH* of the precipitate was further increased by addition of a base (ammonium hydroxide 0.1 M), the nanoparticles redispersed spontaneously, yielding a concentrated solution of individually coated particles. These particles are dubbed PAA$_{2K}$–γ-Fe$_2$O$_3$ in the sequel of the paper [33-35]. From the hydrodynamic diameters of the bare and PAA$_{2K}$-coated particles $D_H^{bare}$ and $D_H^{coated}$ (Table I), the layer thickness was estimated at 3 ± 1 nm (Fig. 1b). The dispersions were further dialyzed against DI-water at *pH* 8 (MWCO 10000 g mol$^{-1}$) to remove unreacted PAA$_{2K}$ chains and the salt in excess. The chemicals used in this work, including poly(acrylic) chains, nitric acid, sodium and ammonium hydroxide and methylene blue were purchased from Aldrich and used without purification. The experimental set-up used for static and dynamic light scattering, transmission electron microscopy, refractometry, thermogravimetric analysis and UV-visible absorbance are described in Supporting Information.

|  | iron oxide nanoparticle | | |
|---|---|---|---|
| $D_0^{VSM}$ (nm) | 6.7 | 8.3 | 10.7 |
| $s^{VSM}$ | 0.21 | 0.26 | 0.33 |
| $D_0^{TEM}$ (nm) | 6.8 | 9.3 | 13.2 |
| $s^{TEM}$ | 0.18 | 0.18 | 0.23 |
| $M_W^{Part}$ (g mol$^{-1}$) | 1.3×10$^6$ | 5.4×10$^6$ | 12×10$^6$ |
| $D_H^{bare}$ (nm) | 13 | 24 | 27 |
| $D_H^{coated}$ (nm) | 18 | 31 | 35 |

**Table I**: Characteristics of the particles used in this work. $D_0^{VSM}$ and $D_0^{TEM}$ are the median diameters of the bare particles determined by Vibrating Sample Magnetometry (VSM) and Transmission Electron Microscopy (TEM), respectively. $s^{VSM}$ and $s^{TEM}$ are the corresponding polydispersities. $M_W^{Part}$ denotes the molecular-weight of the bare particles derived from static light scattering (SLS) experiments. $D_H^{bare}$ and $D_H^{coated}$ are the hydrodynamic diameters of the bare and coated particles, as determined by dynamic light scattering (DLS).

# 3. Results and Discussion

## 3.1 Stability and resilience of the poly (acrylic acid) coating

Dynamic light scattering experiments were conducted to evaluate the stability of $PAA_{2K}$–$\gamma$-$Fe_2O_3$ particles in various solvents, including brines, physiological and cell culture media (SI-5 and Refs. [36, 37]). In contrast to the bare or citrate-coated particles that rapidly precipitate, $PAA_{2K}$–$\gamma$-$Fe_2O_3$ particles were shown to be stable in such media over long periods of time (> years). To demonstrate the resilient anchoring of the organic adlayer at the nanoparticle surface, the hydrodynamic diameter $D_H$ and electrophoretic mobility $\mu_E$ of the particles were measured as a function of the $pH$. Starting at $pH$ 8, nitric acid ($HNO_3$, 0.1 M) was added progressively to the dispersion down to $pH$ 2. The $pH$ was then increased up to $pH$ 10 by addition of ammonium hydroxide (0.1 M). Between two successive additions of acid or base, $D_H$ and $\mu_E$ were determined. The results for the nanoparticles at $D_0^{VSM} = 6.7$, 8.3 and 10.7 nm are shown in Figs. 2 (upper panels). There, the $PAA_{2K}$–$\gamma$-$Fe_2O_3$ particles remained stable from $pH$ 10 down to $pH$ 3.5. It should be noted that $D_H$ decreased by 3 nm for the three samples investigated, with a minimum located around $pH$ 5.5. This result suggests that as the ionization of the chains decreases, the electrostatic interactions between monomers diminish, resulting in a collapse of the $PAA_{2K}$ brush. Below $pH$ 3.5, nanoparticles agglomerated and the dispersion underwent a rapid and macroscopic precipitation. The steric inter-particle interactions mediated by the non-charged $PAA_{2K}$ are not sufficient to counterbalance the van der Waals attractive interactions [38].

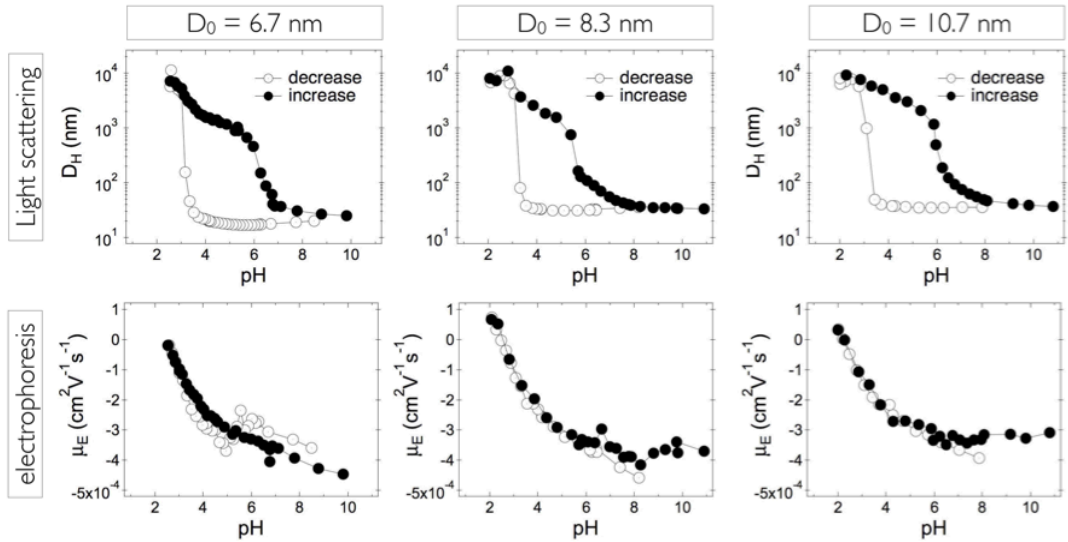

**Figure 2:** *Upper panels* - Hydrodynamic diameters measured on a 6.7, 8.3 and 10.7 nm $PAA_{2K}$–$\gamma$-$Fe_2O_3$ dispersion by dynamic light scattering during $pH$ loop between 2 and 12. *Lower panels* - Electrophoretic mobility $\mu_E$ measured on the same dispersions under the same conditions. The empty and close symbols correspond to $pH$ decrease and increase respectively.

Below $pH$ 3.5, the values of the hydrodynamic diameters obtained from the Malvern instrument are large, and indicative only. Starting from this precipitated state, the $pH$ was then progressively increased by addition of NH$_4$OH under constant stirring. The hydrodynamic diameter started to decrease, exhibited a jump around $pH$ 6 and at the final $pH$, the aggregates redispersed and the particles reached their initial sizes. The hysteretic loops in Figs. 2 are due to kinetic effects in the ionization processes of the carboxylate functions, some of these groups remaining inaccessible in the aggregated state.

The electrophoretic mobility of PAA$_{2K}$–γ-Fe$_2$O$_3$ nanoparticles was also evaluated during the $pH$ loops. The negative $\mu_E$ -values found at $pH$ 8 ($\mu_E \sim -4 \times 10^{-4}\ cm^2 V^{-1} s^{-1}$) corresponding to a zeta potential $\zeta_P \sim -50\ mV$ result from the charged carboxylates located in the polyelectrolyte brush [39]. With decreasing $pH$, $\mu_E$ first increase slowly up to $pH$ 5, and then faster in the lowest range (Figs. 2, lower panels). The aggregation threshold at $pH$ 3 is associated to a mobility of $\mu_E \sim -1.0 \times 10^{-4}\ cm^2 V^{-1} s^{-1}$ for the three samples ($\zeta_P \sim -15\ mV$). Contrary to the size measurements, the electrophoretic mobility exhibited no hysteresis as a function the $pH$. Note also that at low $pH$, the uncomplexed sites at the surface of the 8.3 and 10.7 nm particles become positive, resulting in a slightly positive mobility for the micrometric clusters. From the above measurements, it is concluded that the initial and final states of nanoparticles are equivalent in terms of size and of charge. Remarkably, even in conditions where they are fully protonated, the polymers do not desorb. These results confirm the existence of a highly resilient PAA adlayer onto the γ-Fe$_2$O$_3$ nanoparticles, a property that is crucial for applications in pollutant removal.

## 3.2 Number of poly (acrylic acid) chains per particle

To evaluate the particle anti-pollutant efficiency, the number of adsorbed polymers per particle, noted $N_{ads}^{PAA_{2K}}$ in the following has first to be derived. To this aim, light scattering and thermogravimetric analysis experiments were carried out. Figs. 3 (upper panels) displays the Rayleigh ratios $\mathcal{R}(c)$ for the 6.7, 8.3 and 10.7 nm PAA$_{2K}$-coated particles in the range $c = 0 \ldots 0.4$ wt. %. The scattered intensity is compared to that of the bare particles. For bare and coated non-interacting particles [38], $\mathcal{R}(c)$ varies linearly with $c$ according to:

$$\mathcal{R}(c) = K M_W^{app} c$$

(1)

where $M_W^{app}$ is the apparent weight-average molecular weight of the scattering entities, $K = 4\pi n^2 (dn/dc)^2 / N_A \lambda^4$ is the scattering contrast, $N_A$ is the Avogadro number and $dn/dc$ is the refractive index increment. These increments were determined for the bare particles and for the polymers and found at 0.225 and 0.146 ml g$^{-1}$, respectively. The Rayleigh ratios measured as a function of the concentration yielded molecular weights $M_W^{Part} = 1.3 \times 10^6, 5.4 \times 10^6$ and $12 \times 10^6$ g mol$^{-1}$ for the different oxide particles (Table I). For the PAA$_{2K}$-coated particles, $M_W^{app}$ takes into account the molecular weight of the magnetic core and that of the coating. The departure from linearity observed for the 10.7 nm sample is interpreted as resulting from the repulsive interactions between particles [38]. The slopes obtained for the bare and coated nanoparticles can be utilized to retrieve structural parameters, such as the number of chains per particle. The ratio between the slopes for a given sample reads [34, 40]:

$$\frac{\mathcal{R}_{coated}(c)}{\mathcal{R}_{bare}(c)} = \frac{K_{coated}}{K_{bare}}\left(1 + N_{ads}^{PAA_{2K}}\frac{M_W^{Pol}}{M_W^{Part}}\right)^2$$

(2)

where $K_{coated}$ and $K_{bare}$ are the scattering contrast for the coated and bare nanoparticles. The quantity $K_{coated}$ was derived assuming that for the coated particles $dn/dc$ is the weighted sum of the increments of each component [34]. The ratios between the linear dependencies are 1.59, 1.18 and 1, respectively. From these values, the numbers of polymers per particle $N_{ads}^{PAA_{2K}}$ are estimated. One gets $N_{ads}^{PAA_{2K}} = 200$, 430 and 260 (± 20%) for the 6.7, 8.3 and 10.7 nm batches. For the 10.7 nm sample, the molecular weight of the particle is much larger than that of the organic adlayer, and $N_{ads}^{PAA_{2K}}$ cannot be derived with accuracy. The $N_{ads}^{PAA_{2K}}$ −value of 260 polymers per particle hence must be taken with caution. Thermogravimetric analysis was performed using a TGA 500 (TA Instruments©) on $c$ = 1 wt. % PAA$_{2K}$–γ-Fe$_2$O$_3$ dispersions. With TGA, the number of polymer per particle were found at $N_{ads}^{PAA_{2K}} = 280, 600$ and 810 (± 10%), in fair agreement with the previous determination (Table II). In the sequel of the paper, these values will be considered. They correspond to a density of approximately 2 polymers per nm$^2$. Note that within the experimental errors, these densities are independent on the particle size.

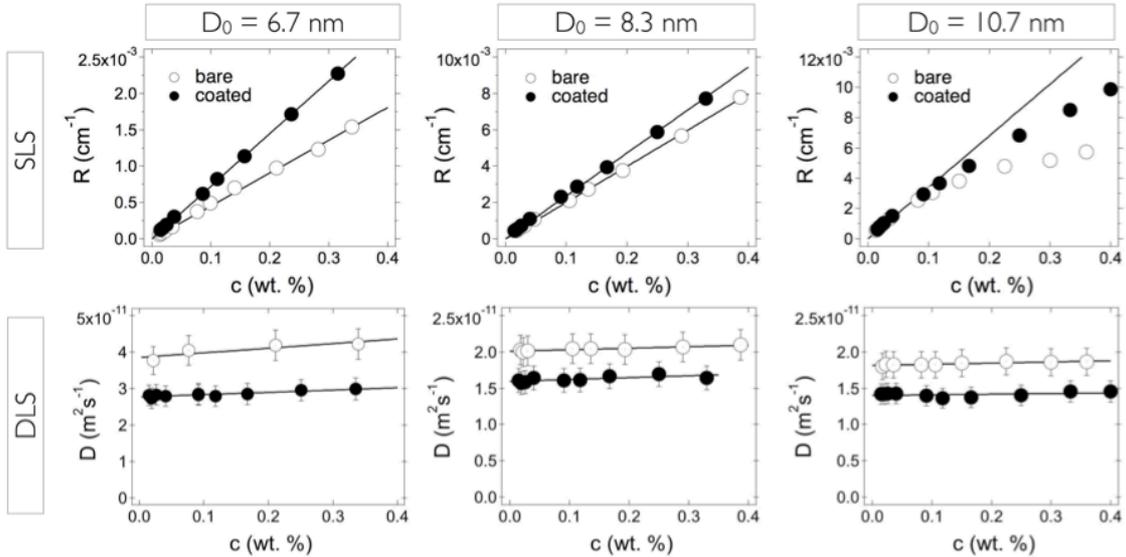

**Figure 3:** *Upper panels* - Rayleigh ratios measured by static light scattering for bare and PAA$_{2K}$-coated particles ($D_0^{VSM}$ = 6.7, 8.3 and 10.7 nm) as a function of concentration. The straight lines are from Eq. 1. *Lower panels* - Diffusion coefficients *versus* concentration obtained on the same iron oxide dispersions. The extrapolation of the coefficient at zero-concentration allows determining the hydrodynamic diameters of the scatterers.

In Figs. 3 (lower panels), the collective diffusion coefficient $D(c)$ obtained by dynamic light scattering is shown *versus* iron oxide concentration for the coated and uncoated samples. As anticipated form dilute solutions, a linear dependence is observed [38]. From the diffusion coefficient extrapolated at zero-concentration, the hydrodynamic diameter of the colloids was retrieved using the Stokes-Einstein relation, $D_H = k_B T / 3\pi \eta_S D$, where $k_B$ is the Boltzmann constant, $T$ the temperature ($T$ = 298 K) and $\eta_S$ the solvent viscosity ($\eta_S = 0.89 \times 10^{-3}$ Pa s for water). The hydrodynamic diameters for the bare and coated 6.7 nm PAA$_{2K}$–γ-Fe$_2$O$_3$ particles are $D_H$ = 13 nm and $D_H$ = 18 nm respectively, indicating a layer thickness of $h$ = 2.5 nm (Fig. 1b). Data for the larger particles are given in Table II.

### 3.3 Electrostatic charge density

In a further step to assess the adsorption properties, the number densities of chargeable groups, *i.e.* of carboxylates need to be determined. To address this question, acid-base titrations were carried out on the different iron oxide batches. The curves of Figs. 4a-c (left scale) display the *pH*-evolution of dispersions upon addition of sodium hydroxide (NaOH 0.036 M). The curves on the right scale (in blue) represent the derivative of the *pH* with respect to the number of moles of base added, $n_{NaOH}$. In the function $dpH/dn_{NaOH}$ $vs$ $n_{NaOH}$, three maxima are observed and indicated by vertical dotted lines. Located at *pH* 4.4 ± 0.1, 8.3 ± 0.2 and 10.5 ± 0.1, the maxima correspond to equivalence points in the acid-base titration. The first equivalence point relates to the titration of the protons arising from the hydrochloric acid used to lower the *pH*. The second equivalence point at *pH* 8.3 is associated with the titration of the carboxylates groups at the particle surfaces, whereas the third equivalence results from the de-protonation of the ammonium counterions. From the second equivalence, two quantities are retrieved: the $pK_a$ of the adsorbed PAA$_{2K}$ (which is defined as the value of the *pH* at the half-equivalence) and the number of moles of hydroxide necessary to protonate all chargeable groups present. Here one finds $pK_a$ = 6.73 ± 0.15, a value that is slightly higher than that of the single polymers, $pK_a$ = 6.14. As suggested by Laguecir *et al.* from experiments and simulations [41], the crowding of the charges at the surface increases the electrostatic interactions, which in turn delay the ionization of the chains with increasing *pH*. Similar $pK_a$ retardance effects were observed with poly(acrylic acid) of high molecular weights (> 100 000 g mol$^{-1}$). From the numbers of carboxylate groups titrated, the average density of carboxylic acid groups $n^{COO^-}$ can be obtained. The inset of Fig. 4a shows the density as a function of the magnetic diameter $D_0^{VSM}$, and reveals a size independent value at $n^{COO^-}$ = 26 ± 4 nm$^{-2}$. Passing from 6.7 nm to 10.7 nm, the number of chargeable groups is hence multiplied by 4.2, from -3040$e$ to -12720$e$. In conclusion of this part, the Precipitation-Dispersion method ensures a uniform and dense coating of the particles, with an elevated and constant density of chains and charges. Combining now the values of the polymer and charge densities, the proportion of adsorbed *versus* uncomplexed monomers per chain can be calculated. As shown in Table II, the proportion of carboxylates dangling in the solvent are comprised between 49% and 72%, the degree of polymerization of the initial chain being 22. In average, 9 ± 3 carboxylates per chain are adsorbed at the surface of the particles, complexing the FeOH$_2^+$ present at low *pH* [42]. Occurring via a multi-site binding mechanism, the electrostatic coupling leads to a solid and resilient anchoring of the PAA$_{2K}$ adlayer. This feature is a prerequisite for the development of applications.

| | iron oxide nanoparticle | | | technique |
|---|---|---|---|---|
| $D_0^{VSM}$ (nm) | 6.7 | 8.3 | 10.7 | VSM |
| $N_{ads}^{PAA_{2K}}$ | 200 | 430 | 260 | SLS |
| $N_{ads}^{PAA_{2K}}$ | 280 | 600 | 810 | TGA |
| $n_{ads}^{PAA_{2K}}$ (nm$^{-2}$) | 2.0 | 2.2 | 1.5 | TGA |
| $n^{COO^-}$ (nm$^{-2}$) | 21 | 30 | 23 | titration |
| percentage of chargeable $COO^-$ | 49% | 62% | 72% | TGA/titration |

**Table II**: List of characteristics of particles used in this work. $D_0^{VSM}$ is the median diameter of the bare particles determined by vibrating sample magnetometry (VSM). The number of PAA$_{2K}$ adsorbed polymers noted $N_{ads}^{PAA_{2K}}$ was determined by static light scattering (SLS) and thermogravimetric analysis (TGA). The uncertainties in the determination of $N_{ads}^{PAA_{2K}}$ are 20% for SLS and 10% for TGA. $n_{ads}^{PAA_{2K}}$ denotes the number density of polymers at the surface, as determined by TGA, and $n^{COO^-}$ the surface density of chargeable carboxylate groups available. The last line gives the proportion of chargeable carboxylate monomers per chain. The degree of polymerization of PANa$_{2.1K}$ is 22.

## 3.4 Methylene blue/nanoparticle interactions

To investigate the trapping ability of the particles, the adsorption of methylene blue (MB) was investigated. Methylene blue is an interesting model of organic pollutant because of its cationic charge and its UV-Vis absorbance properties. MB exhibits a strong peak at $\lambda = 664\ nm$, its extinction coefficient being $9\times10^4\ L\ mol^{-1}cm^{-1}$. More importantly, the peak does not overlap with the broad absorption band of iron oxide (Fig. SI-8). Recently MB was tested against magnetite composites [5], alginate beads [7, 17] and low-cost adsorbents [4], and the adsorption isotherms of MB molecules were derived. In the present study, the amount of adsorbed dyes on PAA$_{2K}$–γ-Fe$_2$O$_3$ was determined as a function of the equilibrium concentration in the supernatant. Practically, 500 μL of a 6.7 nm dispersion at 0.01, 0.05 and 0.1 wt. % were mixed with 500 μL of MB solutions with concentrations between 5 μM and 8 mM. The pristine solutions were adjusted to pH 8.5, ensuring a complete ionization of the carboxylate functions. Mixtures were stirred, centrifuged, and sedimented onto a strong ferrite magnet for 24 hours. The MB concentration in the supernatant was measured by UV-visible spectrometry (Variant spectrophotometer Cary 50 Scan). The calibration curve and the UV-Vis spectra for methylene blue and iron oxide dispersion are reported in Section SI-8.

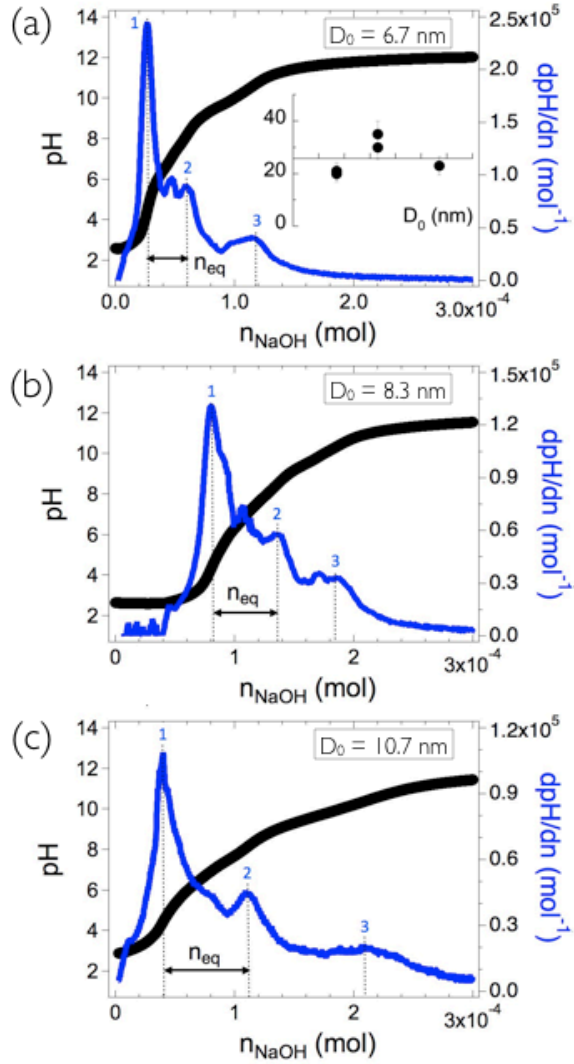

**Figure 4**: Acid-base titrations of PAA$_{2K}$–γ-Fe$_2$O$_3$ dispersions by sodium hydroxide 0.036 M solution under nitrogen atmosphere for $D_0^{VSM}$ = 6.7 nm (a), 8.3 nm (b) and 10.7 nm (c). The closed circles (left scale) represents the $pH$ measured as a function $n_{NaOH}$. The blue line (right scale) is the derivative of the $pH$ with respect to $n_{NaOH}$. $n_{eq}$ stands for the number of moles needed to titrate the carboxylate groups. Inset: number density of carboxylate groups at the nanoparticle surface for the three iron oxide nanoparticles investigated. The horizontal line depicts the average value at $n^{COO^-} = 26 \pm 4$ nm$^{-2}$.

Fig. 5 displays the adsorption isotherm of MB towards 6.7 nm PAA$_{2K}$-coated iron oxide. The amount of adsorbed dyes $q_e$ increases rapidly as a function of $c_e$, the concentration of MB molecules in the supernatant, and then levels off at a plateau value around 2.5 mmol g$^{-1}$. The continuous line in the figure was calculated assuming a Langmuir-type adsorption of the form:

$$q_e(c_e) = q_{max} \frac{Kc_e}{1 + Kc_e}$$

(3)

where $q_{max} = 2.6\ mmol\ g^{-1}$ is the saturation value of the MB adsorption and $K = 5160\ L\ mol^{-1}$ is the binding constant. Eq. 3 assumes the existence of a monolayer adsorbed on the surface and for which the adsorption sites are equivalent and uncorrelated. Translated in mass, the saturation value of $q_{max} = 2.6\ mmol\ g^{-1}$ corresponds to 830 mg g$^{-1}$, that is 4 times larger than the results obtained by Mak et al. (200 mg g$^{-1}$) on a similar system [5]. The difference here can be explained by the higher density of carboxylate functions at the surface of nanoparticles. Adsorption amounts of the order of 830 mg g$^{-1}$ are among the largest measured with BM dyes. There are close from the commercial activated carbon (980 mg g$^{-1}$) or from bio-adsorbents made of polymer-modified biomass of baker's yeast (870 mg g$^{-1}$). PAA$_{2K}$-coated iron oxides are three times more efficient pollutant removals than natural materials and industrial solid wastes [4].

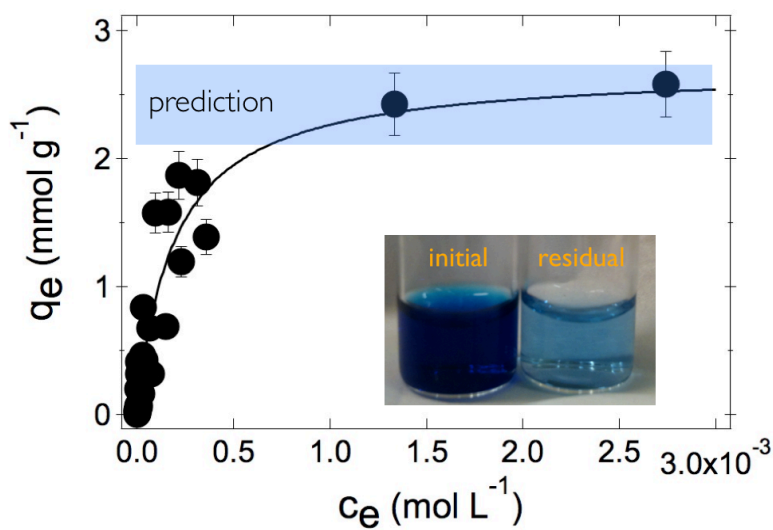

**Figure 5:** Amount of methylene blue adsorbed on the 6.7 nm particles as a function of the dye concentration in the supernatant. The continuous line results from best-fit calculations assuming a Langmuir-type isotherm (Eq. 3). The blue shaded area corresponds to the adsorption at saturation estimated from the number density of chargeable carboxylates at the particle surface. Inset: images of vials containing methylene before ("initial") and after ("residual") interactions with the PAA$_{2K}$-coated nanoparticles.

From the number of carboxylates $n^{COO^-}$ made available at the nanoparticle surface (Table II), the maximum MB removal efficiency can be evaluated, and compared to the above data. Assuming a 1:1 electrostatic-driven complexation, 1 gram of particles corresponds to 0.75 ± 0.08 g of MB molecules, and to a predicted $q_{max}$-value of $2.4 \pm 0.3\ mmol\ g^{-1}$. This value, together with the error bars is indicated in Fig. 5 as the shaded blue area. It is in excellent agreement with the adsorption isotherm at saturation, suggesting that all available carboxylate are participating to the complexation and to the adsorption of MB molecules.

# 4. Conclusion

In view of applications in the field of pollutant remediation, monodisperse poly(acrylic acid) coated iron oxide with diameters 6.7, 8.3 and 10.7 nm were synthesized. With iron oxide as magnetic cores, the separation of the pollutant/nanoparticles complexes is facilitated by the application of an external magnetic field. As compared to uncoated or citrate-coated particles that rapidly precipitate in complex environments, $PAA_{2K}$–$\gamma$-$Fe_2O_3$ particles are stable over long periods of time (> years) [37, 43]. pH loops between pH 2 and 12 applied to the dispersions show that even in conditions where the polyanions are protonated, the polymers do not desorb from the particle surface, confirming the remarkable resilience of the adlayer. In its ionized state ($pH > 6$), the thickness of the adlayer was estimated at 2.5 – 3 nm. In the present paper, the density of carboxylate moieties forming this layer was determined quantitatively. Techniques used were a combination of static and dynamic light scattering, thermogravimetry analysis, acid-base titration and UV-visible absorbance. The density of charged groups was found to be particle size-independent, of the order of $26 \pm 4 \text{ nm}^{-2}$. For a 10 nm particle, this value corresponds to a nominal charge of -8200$e$, *i.e.* much larger than that of the bare particles or of particles coated with low molecular weight ligands [44]. A further advantage of poly(acrylic acid) is that it imparts an electrosteric repulsion to the particles [36].

Among the methods available for cleaning wastewater, additive-based adsorption is used extensively. Low-cost adsorbents made of agricultural and industrial solid wastes, but also of natural nanomaterials have been developed recently, and tested against several dyes molecules and model systems. The performances of these new adsorbents remain however limited. The adsorption isotherm of methylene blue with respect to the anionic coated particles was measured and revealed an unprecedented adsorption at saturation, typically 830 mg g$^{-1}$. This value is larger than those of these new adsorbents mentioned previously, and close to the commercial activated carbon (980 mg g$^{-1}$). Moreover, thanks to the high specific surface, the process of capture and adsorption lasts a few minutes, *i.e.* much faster than with mesoporous beads [21, 45] and hydrogels [7, 25] for which the diffusion into the pores is the limiting process. The present nanotechnology offers an efficient platform with further optimization, in terms of sizes of the different components, and leading to cost-effective pollutant remediation systems.

**Acknowledgements**: Gaelle Charron is acknowledged for her critical reading of the manuscript, and for fruitful discussions. This research was supported in part by the Agence Nationale de la Recherche under the contract ANR-09-NANO-P200-36.

**Electronic Supplementary Material:**
The Supporting Information section describes the characterization techniques used in this work. SI-1: Transmission electron microscopy (TEM); SI-2: Electron beam microdiffraction; SI-3: Vibrating sample magnetometry (VSM); SI-4 – Thermogravimetric Analysis (TGA); SI-5 – Stability of $PAA_{2K}$-coated nanoparticles in cell culture media; SI-6 – Stability and pH loops; SI-7 – Static (SLS) and Dynamic (DLS) Light Scattering, refractometry; SI-8 – UV-Visible Absorbance. Supplementary material is available in the online version of this article at http://dx.doi.org/10.1007/s12274-***-****-* (automatically inserted by the publisher).

# Reference

[1]   S. Behrens, Nanoscale 3 (2011) 877.
[2]   E. Navarro, A. Baun, R. Behra, N.B. Hartmann, J. Filser, A.J. Miao, A. Quigg, P.H. Santschi, L. Sigg, Ecotoxicology 17 (2008) 372.
[3]   H. Wang, Q.-W. Chen, J. Chen, B.-X. Yu, X.-Y. Hu, Nanoscale 3 (2011) 4600.
[4]   M. Rafatullah, O. Sulaiman, R. Hashim, A. Ahmad, Journal of Hazardous Materials 177 (2010) 70.
[5]   S.Y. Mak, D.H. Chen, Dyes and Pigments 61 (2004) 93.
[6]   L.S. Zhong, J.S. Hu, H.P. Liang, A.M. Cao, W.G. Song, L.J. Wan, Advanced Materials 18 (2006) 2426.
[7]   V. Rocher, J.M. Siaugue, V. Cabuil, A. Bee, Water Research 42 (2008) 1290.
[8]   S.H. Huang, D.H. Chen, Journal of Hazardous Materials 163 (2009) 174.
[9]   X.L. Wang, L.Z. Zhou, Y.J. Ma, X. Li, H.C. Gu, Nano Research 2 (2009) 365.
[10]  D. Zhang, S. Wei, C. Kaila, X. Su, J. Wu, A.B. Karki, D.P. Young, Z. Guo, Nanoscale 2 (2010) 917.
[11]  H. Chen, P.K. Chu, J. He, T. Hu, M. Yang, Journal of Colloid and Interface Science 359 (2011) 68.
[12]  B.S. Inbaraj, B.H. Chen, Bioresource Technology 102 (2011) 8868.
[13]  H.M. Sun, L.Y. Cao, L.H. Lu, Nano Research 4 (2011) 550.
[14]  F. Yu, J. Chen, L. Chen, J. Huai, W. Gong, Z. Yuan, J. Wang, J. Ma, Journal of Colloid and Interface Science 378 (2012) 175.
[15]  J.-S. Xu, Y.-J. Zhu, Journal of Colloid and Interface Science 385 (2012) 58.
[16]  I. Ursachi, A. Stancu, A. Vasile, Journal of Colloid and Interface Science 377 (2012) 184.
[17]  L.H. Ai, M. Li, L. Li, Journal of Chemical and Engineering Data 56 (2011) 3475.
[18]  W. Luo, L.H. Zhu, N. Wang, H.Q. Tang, M.J. Cao, Y.B. She, Environmental Science & Technology 44 (2010) 1786.
[19]  K.P. Singh, S. Gupta, A.K. Singh, S. Sinha, Chemical Engineering Journal 165 (2010) 151.
[20]  P.I. Girginova, A.L. Daniel-Da-Silva, C.B. Lopes, P. Figueira, M. Otero, V.S. Amaral, E. Pereira, T. Trindade, Journal of Colloid and Interface Science 345 (2010) 234.
[21]  C. Wang, S.Y. Tao, W. Wei, C.G. Meng, F.Y. Liu, M. Han, Journal of Materials Chemistry 20 (2010) 4635.
[22]  O. Ozay, S. Ekici, Y. Baran, S. Kubilay, N. Aktas, N. Sahiner, Desalination 260 (2010) 57.
[23]  C.L. Chen, X.K. Wang, M. Nagatsu, Environmental Science & Technology 43 (2009) 2362.

[24] J. Hu, D.D. Shao, C.L. Chen, G.D. Sheng, J.X. Li, X.K. Wang, M. Nagatsu, Journal of Physical Chemistry B 114 (2010) 6779.
[25] V. Rocher, A. Bee, J.M. Siaugue, V. Cabuil, Journal of Hazardous Materials 178 (2010) 434.
[26] A.Z.M. Badruddoza, G.S.S. Hazel, K. Hidajat, M.S. Uddin, Colloids and Surfaces a-Physicochemical and Engineering Aspects 367 (2010) 85.
[27] S.R. Shirsath, A.P. Hage, M. Zhou, S.H. Sonawane, M. Ashokkumar, Desalination 281 (2011) 429.
[28] M. Auffan, L. Decome, J. Rose, T. Orsiere, M. DeMeo, V. Briois, C. Chaneac, L. Olivi, J.-L. Berge-Lefranc, A. Botta, M.R. Wiesner, J.-Y. Bottero, Environmental Science & Technology 40 (2006) 4367
[29] A. Petri-Fink, M. Chastellain, L. Juillerat-Jeanneret, A. Ferrari, H. Hofmann, Biomaterials 26 (2005) 2685.
[30] J.P. Chapel, J.-F. Berret, Current Opinion in Colloid & Interface Science 17 (2012) 97.
[31] R. Massart, E. Dubois, V. Cabuil, E. Hasmonay, J. Magn. Magn. Mat. 149 (1995) 1
[32] E. Dubois, V. Cabuil, F. Boue, R. Perzynski, J. Chem. Phys 111 (1999) 7147
[33] J.-F. Berret, O. Sandre, A. Mauger, Langmuir 23 (2007) 2993.
[34] J.-F. Berret, Macromolecules 40 (2007) 4260.
[35] J. Fresnais, J.-F. Berret, B. Frka-Petesic, O. Sandre, R. Perzynski, Advanced Materials 20 (2008) 3877.
[36] B. Chanteau, J. Fresnais, J.-F. Berret, Langmuir 25 (2009) 9064.
[37] M. Safi, J. Courtois, M. Seigneuret, H. Conjeaud, J.-F. Berret, Biomaterials 32 (2011) 9353.
[38] D.F. Evans, K. Wennerström, The Colloidal Domain. Wiley-VCH, New York, 1999.
[39] M. Ballauff, O. Borisov, Current Opinion in Colloid & Interface Science 11 (2006) 316.
[40] S. Louguet, A.C. Kumar, N. Guidolin, G. Sigaud, E. Duguet, S. Lecommandoux, C. Schatz, Langmuir 27 (2011) 12891.
[41] A. Laguecir, S. Ulrich, J. Labille, N. Fatin-Rouge, S. Stoll, J. Buffle, European Polymer Journal 42 (2006) 1135.
[42] M. Nabavi, O. Spalla, B. Cabane, J. Colloid Interface Sci. 160 (1993) 459
[43] M. Safi, H. Sarrouj, O. Sandre, N. Mignet, J.-F. Berret, Nanotechnology 21 (2010).
[44] I.T. Lucas, S. Durand-Vidal, E. Dubois, J. Chevalet, P. Turq, The Journal of Physical Chemistry C 111 (2007) 18568.
[45] Q.Q. Liu, L. Wang, A.G. Xiao, J.M. Gao, W.B. Ding, H.J. Yu, J. Huo, M. Ericson, Journal of Hazardous Materials 181 (2010) 586.